\documentclass[twocolumn,showpacs,floatfix,prl]{revtex4}

\usepackage[dvips]{epsfig}

\usepackage{float}

\begin{document}

\title{Probing Neutral Majorana Fermion Edge Modes with Charge Transport}

\author{Liang Fu and C.L. Kane}
\affiliation{Dept. of Physics and Astronomy, University of Pennsylvania,
Philadelphia, PA 19104}

\begin{abstract}
We propose two experiments to probe
the Majorana fermion edge states that occur at a junction
between a superconductor and a magnet deposited on the surface of a topological
insulator.  Combining two Majorana fermions into a single Dirac fermion on a magnetic
domain wall allows the neutral Majorana fermions to be probed with charge
transport.  We will discuss a novel interferometer for Majorana
fermions, which probes their $Z_2$ phase.   This setup also allows the
transmission of neutral Majorana fermions through a point contact to be measured.  We
introduce a point contact formed by a superconducting junction and show that
its transmission can be controlled by the phase difference across the junction.
We discuss the feasibility of these experiments using the recently
discovered topological insulator Bi$_2$Se$_3$. \end{abstract}

\maketitle

Majorana fermions have attracted interest in condensed matter
physics because their exotic non-Abelian quantum
statistics\cite{mooreread} form the basis for topological
quantum computation\cite{kitaev,review}.  Potential electronic systems
hosting Majorana
fermions include the $\nu=5/2$ quantum Hall state\cite{mooreread,readgreen},
the $p$-wave superconductor Sr$_2$RuO$_4$\cite{dassarma}, and topological
insulator/superconductor structures\cite{fk1,fk2,Beenakker}.  In the $\nu=5/2$
quantum Hall state, a Majorana bound state is associated with the charge $e/4$
quasiparticle, and gapless chiral Majorana fermions form the neutral sector of
the edge states.   Thanks to the $e/4$ charge, the quasiparticle's non-Abelian statistics
can be probed by measuring charge transport of the edge
states\cite{sarma,stern,bonderson}. Recent experiments have shown evidence for
the quasiparticle charge $e/4$\cite{marcus, heiblum}, and there are now
intense efforts to prove or disprove their non-Abelian nature.

Detecting Majorana fermions in superconductors is more
challenging because they are electrically neutral.  In this work, we propose two
experiments to probe neutral
Majorana fermion edge states predicted in superconductor/magnet/topological
insulator structures\cite{fk1}.   Our basic setup, shown in
Fig. 1, involves a grounded superconductor surrounded by two magnets with opposite
out-of-plane magnetization, which are both deposited on the surface of a topological
insulator.  The magnetic domain wall gives rise to chiral
{\it Dirac} fermions that play the role of ``leads" connecting the
superconductor to the source and drain.
An electron incident from the source splits into two
Majorana fermions which take different paths around the edge of the superconductor
and then recombine before going to
the drain.  We will show the source to drain conductance probes the {\it
interference} of the Majorana fermions, forming a novel ``$Z_2$ interferometer".
In addition, we will show that the
{\it transmission} of Majorana fermions through a ``point contact" formed by a
Josephson junction between two superconductors can be measured, and that the
transmission can be tuned by controlling the phase difference across the
junction.

\begin{figure}
\centerline{ \epsfig{figure=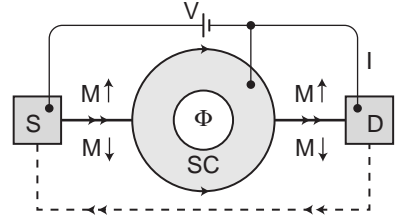,width=2.0in} }
 \caption{An interferometer for Majorana fermions.  Magnetic (M) and
 superconducting (SC) materials are deposited on a topological insulator.  Chiral Majorana
 fermion edge states (denoted by a single arrow) circle the outer boundary of
 the superconductor, and chiral Dirac fermion edge states (denoted by the double
 arrow) are confined to the magnetic domain wall connected to a source (S) and
 drain (D).  A return path between the drain and source is shown with the dashed
 line.   When a voltage is applied to the source electrons are split into two
 Majorana fermions, allowing their $Z_2$ interference phase $\pm 1$ to be
 probed by measuring the current in the drain.}
\end{figure}

A topological insulator\cite{fkm,moore} has gapless surface states that are
topologically protected in the absence of time reversal or
gauge symmetry breaking fields.  Breaking time reversal symmetry either by an
applied magnetic field or by depositing a magnetic material can open an energy
gap leading to a novel surface quantum Hall effect with $\sigma_{xy} = \pm
e^2/2h$\cite{fkm,fukane,zhang}.  Depositing a superconductor on the surface leads, via the proximity effect,
to a surface superconducting state that hosts Majorana fermions\cite{fk1}.
In view of the recent experimental discoveries of topological insulator phases
in Bi$_x$Sb$_{1-x}$\cite{hasan1,hasan2} and Bi$_2$Se$_3$\cite{hasan3}, and the
earlier experimental evidence of good contact between
superconducting Nb and Bi$_x$Sb$_{1-x}$\cite{kasumov}, the
experimental study of these novel gapped phases is now possible.

The superconducting and magnetic phases of the surface states, as well as the
gapless states at interfaces between them, can be
described with the Bogoliubov de Gennes (BdG)
formalism.  The Hamiltonian is $H =
\Psi^\dagger {\cal H} \Psi/2$, where $\Psi = ( \psi_\uparrow, \psi_\downarrow,
\psi_\downarrow^\dagger,-\psi_\uparrow^\dagger)^T$ and
\begin{eqnarray} {\cal H} & =&  \tau^z \left[ v_F \hat{z} \cdot \vec \sigma
\times ( - i  \nabla - e {\bf A} \tau_z)- \mu \right] \nonumber \\ &+& ( \Delta
\tau^+  + \Delta^* \tau^-) + M \sigma_z. \label{hbdg}
\end{eqnarray}
 Here $\psi_\uparrow$ and $\psi_\downarrow$ are electron operators of the
surface states which are Kramers degenerate at ${\bf k}=0$.
 The first line in $\cal H$ describes the free surface states coupled
 to the vector potential $\bf A$.  $\vec \sigma =(\sigma^x,\sigma^y)$
 are Pauli matrices, $v_F$ is the Fermi velocity
 and $\mu$ is the chemical potential.
 $ \Delta \psi_\uparrow^\dagger \psi_\downarrow^\dagger + h.c$ describes the
 superconducting proximity effect.
Spatially uniform $\Delta$ gives a gapped excitation spectrum
$E^s_{\bf k} =\sqrt{(\pm v|{\bf k}| - \mu)^2 + \Delta^2}$.
$M \psi^\dagger \sigma_z \psi$ describes the Zeeman splitting due to the magnet.
Spatially uniform $M$ gives $E^z_{\bf k} =
\sqrt{  v^2 |{\bf k}| ^2 + M^2} \pm \mu$, which is gapped when $ M > \mu$.
The BdG Hamiltonian has particle-hole symmetry, expressed by
$\{\Xi,{\cal H}\}=0$ where the particle-hole operator is
$\Xi \xi = \sigma^y \tau^y \xi^*$.
The eigenstates $\xi_{\pm E}$ with energy $\pm E$ obey $\xi_{-E} = \Xi \xi_E$,
and only the $E \geq 0$ half of the spectrum represents independent excitations.

An interface between two half planes ($y>0$ and $y<0$) with different mass terms
gives rise to gapless 1D domain-wall states.  First consider a
superconductor-magnet interface modeled by $\Delta=\Delta_0 \Theta(y)$
and $M=M_0 \Theta(-y)$. Solving (\ref{hbdg}), we
find one chiral branch of bound states with a
four component wavefunction $\xi_{k}(x,y)$ localized near $y=0$.
$\xi_{k=0}$ has zero
energy and satisfies $\Xi \xi_0 =\xi_0$, which fixes its
phase up to a $\pm$ sign.   Using
$k\cdot p$ theory the eigenstates for small $k$ are
$\xi_{k} (x,y)=\exp(i k x)
\xi_0(y)$ with energy $E(k)=\hbar v_M k$, where
$v_M = v_F  \langle
\xi_0|\tau_z \sigma_y | \xi_0\rangle = v_F \sqrt{1-\mu^2/M_0^2}/(1+\mu^2/\Delta_0^2)$.  These define
Bogoliubov operators $\gamma_{k} = \int dx dy \xi_{k}(x,y)^\dagger
\Psi(x,y)$ which satisfy $\gamma_{k}^\dagger = \gamma_{-k}$.  The
continuum operators $\gamma(x) \sim \int dk \gamma_{k} e^{i k x}$ are
Majorana fields, $\gamma^\dagger(x) = \gamma(x)$ obeying the
low energy Hamiltonian $H = -i \hbar v_M \gamma \partial_x \gamma$.

To model a magnetic domain wall we take $M=M_0 {\rm sgn}(y)$.
We find a gapless branch of chiral edge
states between $\sigma_{xy}=\pm e^2/2h$.  When
expressed in the BdG formalism, {\it two} chiral branches of bound states
with energy $E(k) \sim \hbar v_D k$ appear due to the double counting.
For $E(k)>0$, the two states have the form
$f_{k}\otimes|\tau_z=1\rangle$ and $\Theta f_{-k}\otimes|\tau_z=-1\rangle$.
where
$f_{k}(x,y)$ is a two component wavefunction in the $\sigma_z$ sector
and $\Theta f = \sigma_y f^*$ is the time reversal operator.
These correspond to the electron operators  $c_{k}^\dagger$ and $c_{-k}$
respectively.

To analyze the device in Fig. 1, we employ the BCS mean field theory to
calculate the transport current due to {\it quasiparticles}.  This is justified
because the superconducting order parameter at the surface inherits its phase
from the bulk 3D superconductor, which behaves classically at low temperature.
When the source is biased at a subgap voltage $V \ll \Delta_0$ the
quasiparticles involved are exclusively the gapless Majorana fermion edge
states.

An electron incident from the source can
be transmitted to the drain as an electron, or converted to a hole by an
Andreev process in which charge $2e$ is absorbed into the superconducting
condensate.   Before solving
the general source to drain transmission problem we will show that the behavior at $E=0$
follows from a simple argument.
Scattering at the left
tri-junction, where the incident Dirac fermion meets the superconductor,
 must transform an incident $E=0$ electron
$c_L^\dagger$ into a fermion $\psi$ built from the
Majorana operators $\gamma_1$ and $\gamma_2$.
The arbitrary sign
of $\gamma_{1,2}$ allows us to choose $\psi=\gamma_1 + i \gamma_2$.
Likewise, scattering at the right tri-junction transforms $\psi$ into a
fermion in the right lead.  This must be {\it either}
$c_R^\dagger$ or $c_R$.  A superposition of the two is not allowed because
it is not a fermion operator.
To determine
which occurs, we observe that when the size of the superconductor
shrinks continuously to zero, the left and right lead seamlessly connect to
each other. Adiabatic continuity thus dictates that an incident $E=0$
electron is transmitted as an electron, $c_L^\dagger \rightarrow c_R^\dagger$. When the ring
encloses a quantized flux $\Phi=nh/2e$, this adiabatic argument breaks
down. Instead odd $n$ introduces a branch cut for one of the
Majorana modes, i.e. $\gamma_1 \rightarrow -\gamma_1$.
Thus, when the ring encloses an odd number of
flux quanta, $c_L^\dagger \rightarrow c_R$, and an incident $E=0$ electron is converted to a hole.

To obtain the scattering probabilities at finite energy $0<E \ll \Delta$, we
use the BdG formalism to solve the scattering problem in the limit that the size of
the ring $L$ is much larger than the decay length of the Majorana edge states
into the bulk, which is of order $\max(\hbar v_F/\Delta_0,\hbar v_F/M_0)$.  First consider the
scattering at the left tri-junction. A $2\times 2$ scattering matrix $S(E)$ relates the two
incoming states in the left lead $| \tau_z=\pm 1 \rangle$, which we denote $e$
and $h$ (for electron and a hole), to the two outgoing Majorana
edge states $\xi_1$ and $\xi_2$ on the top and bottom of the ring,
$(\xi_1, \xi_2)^T=S(E)(e, h)^T$.
To simplify the notation, we have used the channel label to denote the amplitude
of the scattering states in the corresponding channel. Particle-hole
symmetry implies that
$S(E)=S^*(-E) \tau_x $.
At $E=0$, this property, along with
unitary $S^\dagger S = 1$, allows $S$ to be chosen as
\begin{equation} S=
\frac{1}{\sqrt 2} \left( \begin{array}{cc} 1 & 1 \\ i & -i \end{array} \right),
\label{S} \end{equation}
so that
$c^\dagger \rightarrow (\gamma_1 - i \gamma_2)/\sqrt2$.  Another solvable limit
is when the BdG Hamiltonian has a mirror symmetry
 ${\cal H}(-y)={\cal M}^{-1} {\cal
H}(y) {\cal M}$ with ${\cal M} = i\sigma_y$. The electron and hole channels are
eigenstates of ${\cal M}$ with eigenvalue $\pm i$, whereas the two Majorana fermion
edge states are interchanged.  This leads to
(\ref{S}) at any energy. To obtain the
exact scattering matrix at $E \neq 0$ for a tri-junction without mirror
symmetry requires solving the $2D$ scattering problem. Here we
assume that Eq.(\ref{S}) is a good approximation of the scattering matrix at
low energy.

Next we study the propagation of the chiral Majorana fermion.
When there is no magnetic flux,  in the semi-classical limit the
wavefunction at energy $E\ll \Delta_0$ can be approximated by $\xi(l,s)=\xi_0(s) \exp( i
k(E) l )$, where $l$ parameterizes the length along the interface, $s$
parameterizes the distance perpendicular to the interface and
$k(E)= E/v_M$. In the presence of a magnetic flux $\Phi=nh/2e$, the
superconducting phase $\phi$ winds by $2\pi n$ around the ring accompanied by a
vector potential $A=\nabla \phi$. It is convenient to choose a gauge in
which the spatial variation of $\phi$ is concentrated near the middle of the upper
semi-circle. Away from this ``scattering region'', the wavefunction is a
free Majorana edge mode as before.  This scattering problem can be solved
with a $U(2)$ gauge transformation that eliminates the spatial variation of
$\phi$ and the nonzero $A$.  The wavefunction of the exact scattering state is
simply the undisturbed wavefunction multiplied by $\exp[i \tau_z \phi (l) /2]$.
We conclude that in the presence of a magnetic flux $\Phi=nh/2e$ the
chiral Majorana edge mode $\gamma_1$ acquires an {\it additional} phase shift
$n \pi$ across the junction.

The scattering amplitude of the ring is found by
composing the scattering matrices:
\begin{eqnarray} \left[
\begin{array}{c} e \\ h \end{array} \right]_R &=& S^{-1} \cdot \left(
\begin{array}{cc} e^{i \pi n + i k l_1 } & 0 \\ 0 & e^{ i k  l_2 } \end{array}
 \right)
 \cdot S
 \left[
 \begin{array}{c}
e  \\ h
\end{array}
\right]_L.
\label{SM}
\end{eqnarray}
The current in the drain when the source is biased at voltage $V$ and the
superconductor and drain are grounded is
\begin{equation}
I =  (-1)^n {e\over h} \int_{0}^{\infty} d E \left[ f(E-eV)- f(E+eV) \right] \cos
\theta(E),
\label{iv1}\end{equation}
where $f$ is the Fermi-Dirac
distribution function and $ \theta  = k ( l_1 - l_2) \equiv  E \delta L /v_M$ is
the relative phase between two paths of different lengths.
Evaluating the integral we find
\begin{equation}
I= (-1)^n {e\over h}\frac{
\pi k_B T  \sin(eV \delta L/v_M )}{\sinh(\pi k_B T \delta L /v_M )}, \;\;\; k_B
T,eV \ll \Delta_0.
\end{equation}
At fixed bias, the current ``oscillates" as a function of the {\it discrete} magnetic
flux $n h/2e$, reflecting the Aharonov Bohm phase for Majorana fermions, which takes values
$\pm 1$.  Our device thus functions as a ``$Z_2$ interferometer" for Majorana
fermions.  The ``visibility" of these oscillations is suppressed below a temperature scale
$k_B T_{\delta L} \equiv \hbar v_M/\delta L$ due to thermal averaging.
In addition, at finite bias voltage the current oscillates as a function of $V$
with a period $2\pi k_B T_{\delta L}/e$ due to the energy dependence of the
relative phase.
That the oscillation persists
to high bias voltages without any damping is due to the absence of
dephasing in our calculation. A similar situation occurs in the electronic Mach-Zehnder
interferometer: the decay of the {magnitude} of interference
oscillation at high bias voltage is attributed to dephasing
processes\cite{heiblum2}. Sources of dephasing in our system include coupling
of Majorana fermions with other degrees of freedom, as well as interactions
between Majorana fermions. Since Majorana fermions are neutral, we expect
environmental coupling is weak. In addition, the lowest order
{\it local} interaction term within the Majorana fermions is $\gamma(x)
\partial_x \gamma(x) \partial_x^2 \gamma(x) \partial_x^3 \gamma(x)$, which
involves spatial derivatives at sixth order and will be strongly suppressed
at low temperature.   Thus there is reason to expect the
low temperature dephasing rate for the Majorana fermion edge states will be
smaller than that of ordinary electrons.

We next study the transmission of Majorana fermions across a
Josephson junction between two superconductors, shown in Fig. 2a.
The junction plays the role of a {\it point
contact} for Majorana fermions and
can be characterized by a scattering matrix relating
incoming and outgoing Majorana modes,
$\gamma_i^{\rm out} = S^{\rm pc}_{ij}(E) \gamma_j^{\rm in}$.
Each superconductor is
connected to a source and drain by chiral electron modes at magnetic domain
walls.  An incident electron from $S_1$ splits into two Majorana modes.
One of the two is scattered by the junction, and has a probably amplitude
$t = S^{\rm pc}_{11}$ of being transmitted and
recombining with its partner before going to $D_1$.  Following the previous procedure, we
calculate the scattering matrix relating an incident fermion at $S_1$
to an outgoing fermion at $D_1$ to obtain the current flowing to $D_1$ when
$S_1$ is at voltage $V$ and the other leads are grounded.
\begin{equation}
 I =  e \int_{0}^{\infty} d E [ f (E-eV)- f(E+eV) ]
 \textrm{Re}[t(E)e^{i\theta(E)}],
\end{equation}
where $\theta(E)$ is the same as in (\ref{iv1}).
At $E=0$ particle-hole symmetry
constrains $S_{pc}$ to be a real $O(2)$ matrix describing the transmission $t =
\cos \delta$ and reflection $r = \sin\delta$ such that
$\gamma_1^{\rm out}+i \gamma_2^{\rm out} = e^{i\delta}(\gamma_1^{\rm in} + i
\gamma_2^{\rm in})$.  The zero bias, zero temperature conductance,
$G = I_{D1}/V_{S1} = t e^2/h$ directly measures the transmission of the neutral
Majorana fermions at the junction.

\begin{figure} \centerline{ \epsfig{figure=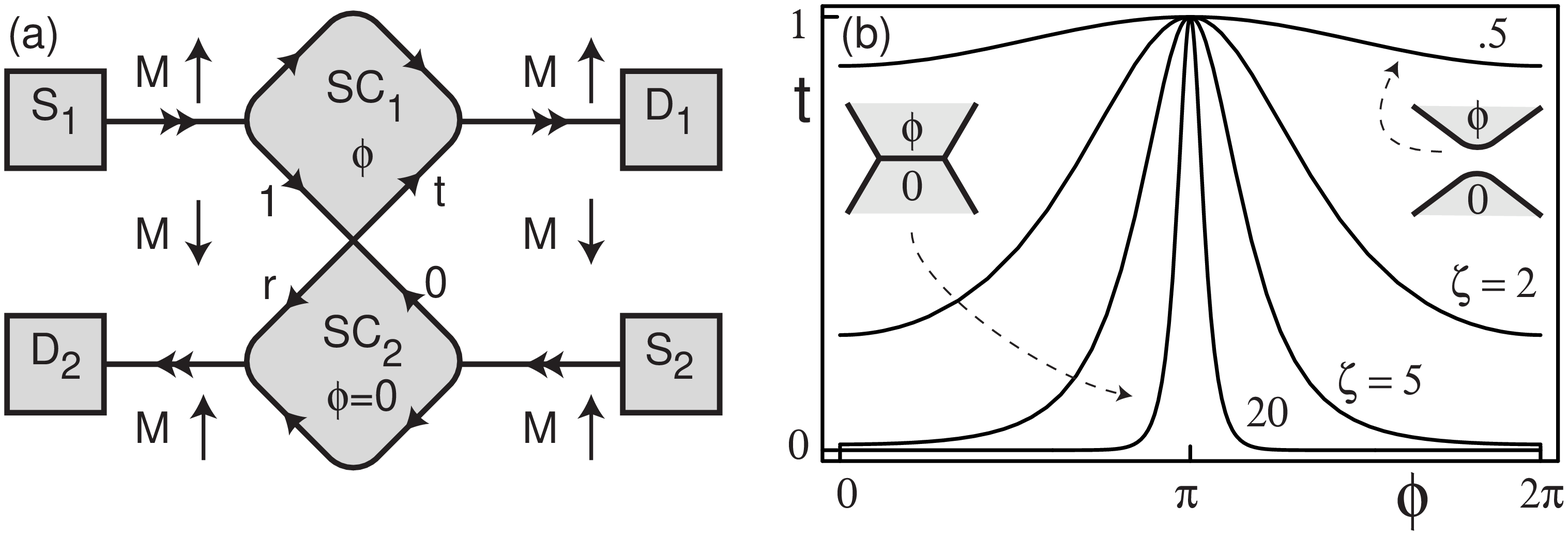,width=3.2in} }
 \caption{(a) A point contact for neutral Majorana fermions characterized by reflection and transmission
 amplitudes $t$ and $r$ formed by a junction between
 two superconductors.  Each superconductor is connected to a source and drain by
 chiral fermions at a magnetic
 domain wall, allowing $t$ to be measured with charge transport.
 (b) Zero energy transmission of the point contact as a function of phase $\phi$ for
 different coupling strengths.  The insets indicate the limits of a weakly coupled point contact (right)
 and a long line junction (left).}
\end{figure}

The transmission amplitude $t$ can be controlled by adjusting the phase difference
$\phi$ of the Josephson junction. $t(\phi)$  depends on the geometry of the
junction. We consider a simple model,
\begin{equation}
H=(\gamma_1, \gamma_2) [ -i  v_M \tau^z \partial_x + \lambda(x) \cos(\phi/2) \tau_y ] (\gamma_1, \gamma_2)^T.
\end{equation}
When $\lambda(x)=\lambda \delta(x)$ and $\lambda/v_M \ll 1$, $H$ describes superconductors
weakly coupled by single electron tunneling at a point\cite{kitaev2, kwon, fk2}.
When $\lambda(x)=\Delta_0$ for $x \in [0, L]$ and $0$ otherwise, $H$ becomes the low
energy theory of a line junction\cite{fk1}.  The transmission amplitude at
$E=0$ in this model is
\begin{equation}
t(\phi) = 1/\cosh[ \zeta \cos(\phi/2) ],
\end{equation}
with $\zeta = \int dx \lambda(x) /2v_M$. Fig. 2b shows $t(\phi)$
for different values of $\zeta$. At
$\phi=\pi$, the transmission is perfect.  This is
guaranteed by gauge invariance.  When
$\phi \rightarrow \phi + 2\pi$ one of the Majorana edge
modes changes sign\cite{fk2} so $r(\phi)= - r(\phi+2\pi)$.  Thus,
$r(\phi)=0$ and $t(\phi)=1$ for some $\phi\in[0,2\pi]$.  For a symmetric
junction this occurs at $\phi=\pi$.

For a weakly coupled point contact (Fig. 2b, right inset), $t(\phi)$ is
energy-independent, but is only weakly
dependent on $\phi$.  For a long line junction, (Fig. 2b, left inset) $t(\phi)$ varies over
a wide range of values between $0$ and $1$, but has a very narrow peak
$\delta \phi \sim \hbar v_M/\Delta_0 L$.  In addition,
near the peak the transmission will be strongly energy dependent due to the
small gap when $\phi\sim \pi$.  It is desirable to engineer the size and
geometry of the Josephson junction in between these two limits, so that
$t(\phi)$ has a well defined peak which can be probed by the low temperature
conductance.

It is worthwhile to compare the superconducting point contact for Majorana fermions
studied here with a point contact in the $\nu= 5/2$ quantum Hall effect.
Our point contact is precisely equivalent to the {\it neutral sector} of the $\nu=5/2$ point
contact, which has been described in terms
of the Ising boundary conformal field theory\cite{ffn}.  For $\nu=5/2$,
however, the physics is dominated by the backscattering of
charge $e/4$ quasiparticles, which is analogous to quantum tunneling {\it vortices}
across the superconductor in our system.  Since the superconducting phase is essentially a
classical variable, this process is strongly suppressed in a superconducting
point contact.  Thus, unlike the $\nu=5/2$ problem, vortex backscattering does
{\it not} lead to a crossover to the weak tunneling limit.

The recently discovered topological insulator Bi$_2$Se$_3$\cite{hasan3,zong}, which has a large bulk
gap $\sim .35$eV is a promising material to probe these states.  Unlike
Bi$_{1-x}$Sb$_x$, its surface states have a small Fermi surface that encloses a single
Dirac point.
Photoemission experiments reveal a Fermi velocity $\hbar v_F \sim .3$eV
nm and a Fermi energy $\mu \sim .3$eV relative to the Dirac point.  The current
materials are unintentionally doped, with the bulk Fermi energy in the conduction
band.  If the material can be compensated either by doping or gating, it is likely
that the surface Fermi energy can be made much closer to the Dirac point.
This is important because achieving the magnetic gapped state
requires a field $M>\mu$.  Moreover, the $k\cdot p$ theory predicts that
the Majorana velocity $v_M$ is suppressed when $\Delta_0 \ll \mu$, reducing
the temperature scale $T_{\delta L}$ required to observe the signature of Majorana fermions.
Our model calculation gives $v_M \sim v_F (\Delta_0/\mu)^2$.
Assuming a superconductor can be found that gives a proximity induced gap
$\Delta_0 \sim .1$meV, we require size $L > \hbar v_F/\Delta_0 \sim 3 \mu$m.
If $\mu \sim 1$meV and $\delta L \sim 1 \mu$m then $T_{\delta L} \sim  30$mK.
$T_{\delta L}$ can be larger if the path difference $\delta L$ can be finely tuned.

To conclude, we have proposed experiments to probe the interference and
transmission of neutral Majorana fermions with charge transport.
We hope they offer a first step towards the more ambitious goal\cite{fk1}
of detecting the non-Abelian statistics of
individual Majorana bound states and using them for quantum
computation.

In a recent preprint, Akhmerov, et al.\cite{beenakker2} independently studied an
interferometer similar to Fig. 1.
We thank Carlo Beenakker for an insightful discussion.
This work was supported by NSF grant
DMR-0605066 and ACS PRF grant 44776-AC10.

\end{document}